\documentclass[a4paper]{mem}
\usepackage{natbib}
\usepackage{graphicx}
\usepackage[a4paper]{hyperref}
\idline{73}{1}
\input epsf
\def\apj{ApJ}
\def\apjs{ApJS}

\def\aj{AJ}
\def\apjl{ApJL}

\def\aap{A\&A}

\def\jcp{J. Comp. Phys.}

\def\s2n{S^{\prime}/N}

\bibpunct{(}{)}{;}{a}{}{,}

\begin{document}
\title{Brown Dwarfs from Turbulent Fragmentation}

\author{
Paolo Padoan\inst{1}, Alexei Kritsuk\inst{1}, Michael, L. Norman\inst{1} \and \AA ke Nordlund\inst{2}\fnmsep
}
\institute{Department of Physics and Center for Astrophysics and Space Sciences,
University of California, San Diego, 9500 Gilman Drive, La Jolla, CA 92093-0424; ppadoan@ucsd.edu, akritsuk@cosmos.ucsd.edu, mnorman@cosmos.ucsd.edu \\
\and Astronomical Observatory / NBIfAFG, Juliane Maries Vej 30, DK-2100, Copenhagen, Denmark; aake@astro.ku.dk
}

\abstract{
The origin of brown dwarfs (BDs) is an important component of the 
theory of star formation, because BDs are approximately as numerous 
as solar mass stars. It has been suggested that BDs originate from 
the gravitational fragmentation of protostellar disks, a very different 
mechanism from the formation of hydrogen burning stars. We propose 
that BDs are instead formed by the process of turbulent fragmentation, 
like more massive stars. In numerical simulations of turbulence and 
star formation we find that gravitationally unstable density peaks 
of BD mass are commonly formed by the turbulent flow. These density 
peaks collapse into BD mass objects with circumstellar disks, like 
more massive protostars. We rely on numerical experiments with very 
large resolution, achieved with adaptive mesh refinement (AMR). The 
turbulence simulation presented here is the first AMR turbulence 
experiment ever attempted and achieves an effective resolution of 
$1024^3$ computational zones. The star formation simulation   
achieves an effective resolution of $(10^6)^3$ computational zones, 
from a cloud size of 5~pc to protostellar disks resolved down to 1~AU. 
}
\authorrunning{Padoan et al.}
\titlerunning{Brown Dwarfs from Turbulent Fragmentation}
   \maketitle

\section{Introduction}

Brown Dwarfs (BDs) are approximately as abundant as solar mass stars 
\citep[e.g.][]{Bejar+2001,Chabrier2002}. 
The typical Jeans' mass in star--forming clouds is of order a solar
mass, approximately two orders of magnitude more massive than a BD. 
This is usually considered the fundamental reason why solar mass 
stars are so common, but the existence of BDs is a challenge to 
this simple explanation. The origin of BDs is apparently an important 
test for the theory of star formation. 

From a different perspective, we could instead assume that the
origin of BDs is quite different from that of more massive stars,
in which case their existence would not necessarily provide 
any constraints to the theory of star formation. This approach is
exemplified in the recent suggestion by \cite{Reipurth+Clarke2001}
that BDs are the result of the gravitational fragmentation of 
protostellar disks, but that work does not draw a definite
line separating BDs from more massive stars. At what mass does 
star formation switch from the standard mechanism to the disk
fragmentation mode? Most likely not exactly at the hydrogen
burning limit of 0.075~m$_{\odot}$, as this limit is defined by 
nuclear physics and can hardly affect the fragmentation process.

There is also no indication of a discontinuity in the stellar
initial mass function (IMF) at approximately 0.075~m$_{\odot}$,
and no evidence of special properties of young BDs relative to
more massive stars, suggesting that perhaps a unique
process can explain the formation of both hydrogen burning stars 
and BDs \citep[e.g.][]{Jayawardhana+2003a,Jayawardhana+2003}. 

The possibility that BDs and more massive stars have a common origin
due to the process of turbulent fragmentation has been recently 
proposed by \cite{Padoan+Nordlund04bd}. In that work, we
show that BDs are predicted by our model for the origin of the 
stellar IMF from turbulent fragmentation \citep{Padoan+Nordlund02IMF},
roughly as frequently as inferred from observations.
Here we briefly summarize our theoretical predictions and present 
new numerical experiments.

\section{Stellar and BD Masses from Turbulent Fragmentation}

The gas density and velocity fields in star--forming clouds are highly
non--linear due to the presence of supersonic turbulence. The kinetic
energy of the turbulence is typically 100 times larger than the gas thermal
energy on the scale of a few pc (the typical rms Mach number is of 
order 10) and the gas is roughly isothermal, so that very large
compressions due to a complex network of interacting shocks cannot be
avoided. Under such conditions the concept of gravitational instability,
based on a comparison between gravitational and thermal energies alone
in a system with mild perturbations, does not apply. Turbulent density 
peaks of any size can be formed in the turbulent flow, independent of 
the Jeans' mass. Peaks that are massive and dense enough (more massive 
their own Jeans' mass) collapse into protostars, while smaller subcritical
ones re-expand into the turbulent flow. This is a process that we call
{\it turbulent fragmentation}\footnote{This term was probably first introduced
by \cite{Kolesnik+Ogulchanskii90}}, to stress the point that stars and BDs,
formed in supersonically turbulent clouds, are not primarily the 
result of gravitational fragmentation.

\begin{figure*}[ht]
\centerline{
\epsfxsize=6.5cm \epsfbox{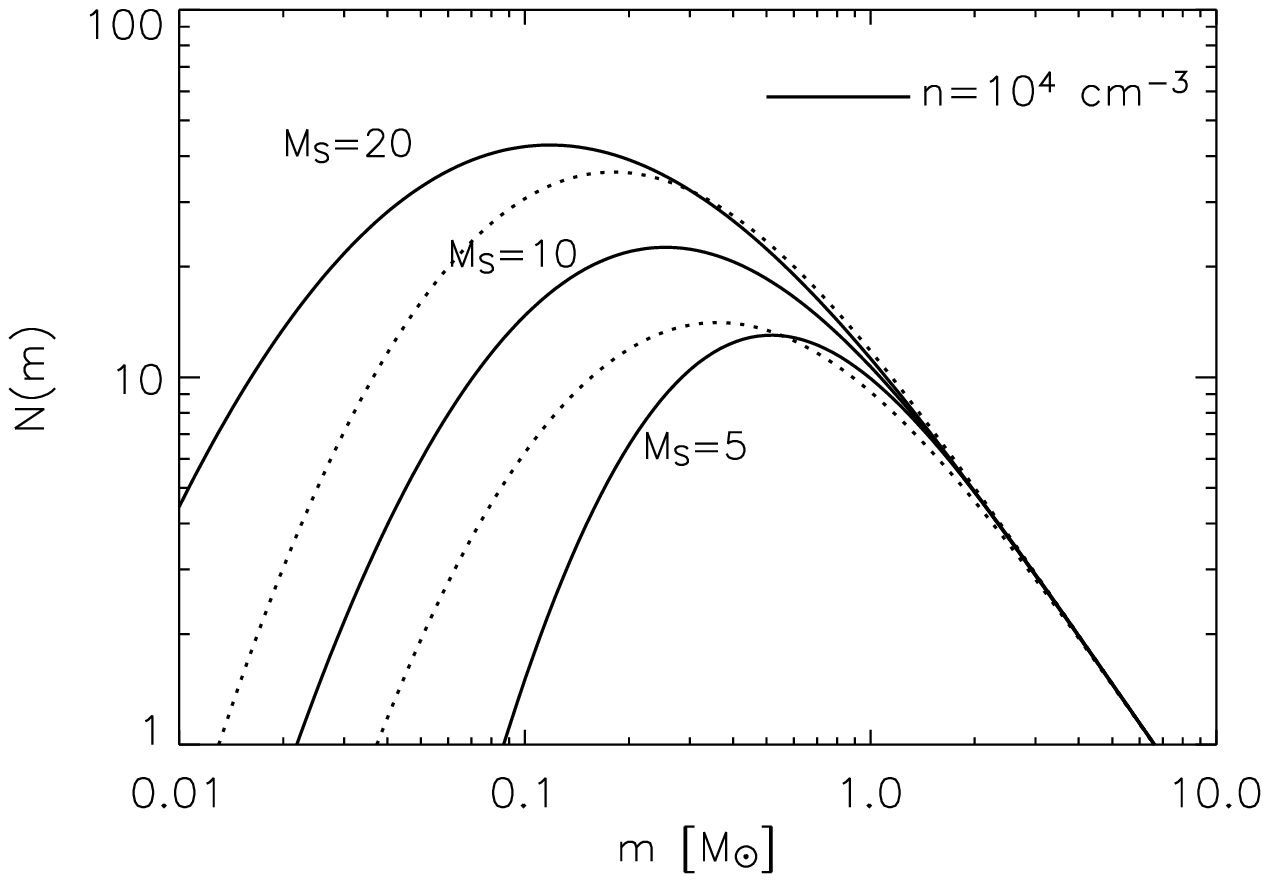}
\epsfxsize=6.5cm \epsfbox{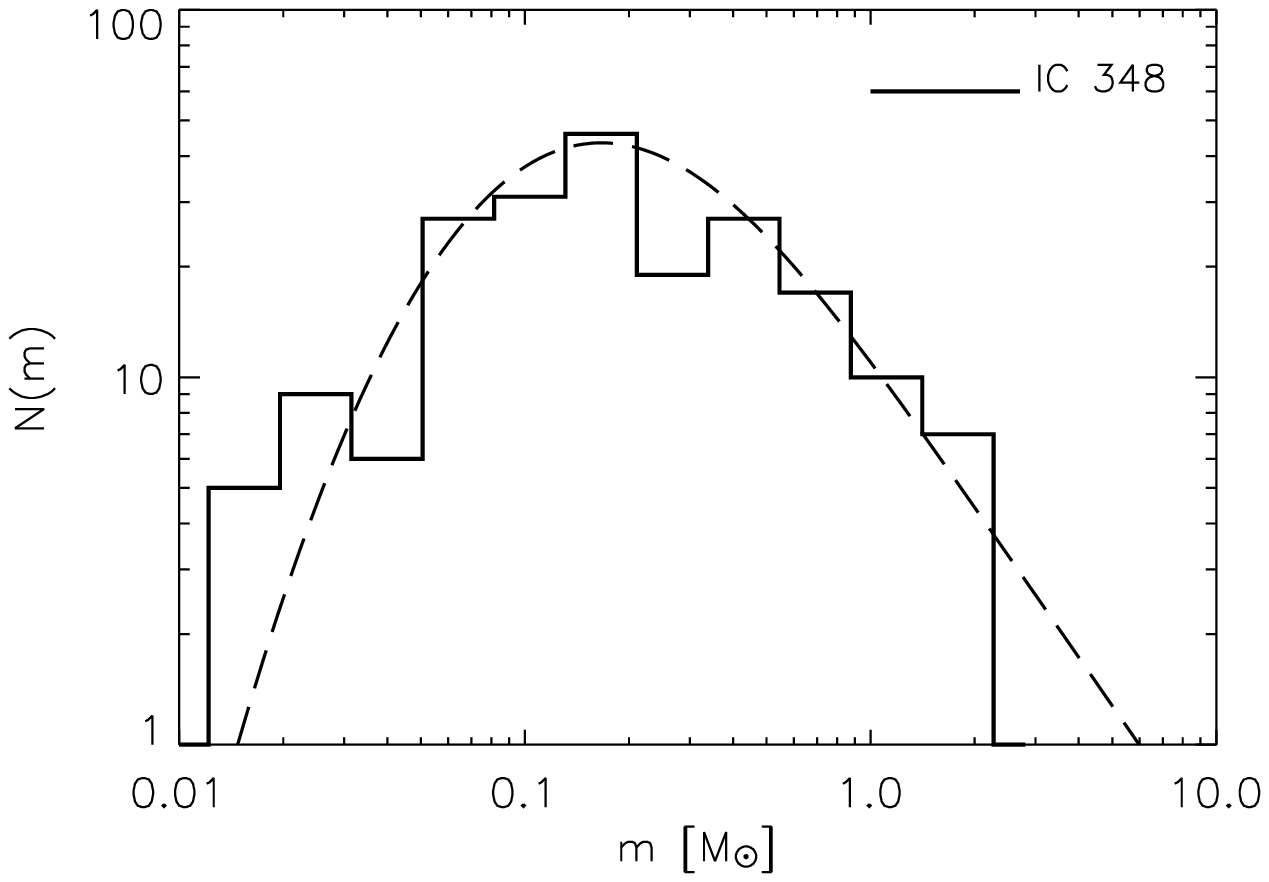}
}
\caption[]{Left panel: Analytical mass distributions computed
for $\langle n\rangle=10^4$~cm$^{-3}$, $T=10$~K and 
for three values of the sonic rms Mach number, $M_{\rm S}=5$, 10 and 20
(solid lines). The dotted lines show the mass distribution for 
$T=10$~K, $M_{\rm S}=10$ and $\langle n\rangle=5\times10^3$~cm$^{-3}$
(lower plot) and $\langle n\rangle=2\times10^4$~cm$^{-3}$ (upper plot).
Right panel: IMF of the cluster IC 348 in Perseus obtained by 
\cite{Luhman+2003} (solid line histogram) and theoretical IMF computed for 
$\langle n\rangle=5\times 10^4$~cm$^{-3}$, $T=10$~K and 
$M_{\rm S}=7$ (dashed line).}
\label{f1}
\end{figure*}

Because density peaks formed by the turbulent flow need to be larger 
than their critical mass to collapse, a necessary condition for
the formation of BDs by supersonic turbulence is the existence of
a finite mass fraction, in the turbulent flow, with density at 
least as high as the critical one for the collapse of a BD mass 
density peak. As the PDF of gas density in supersonic turbulence
is fully determined by the mean density and the rms Mach number of
the turbulent flow \citep{Nordlund+Padoan_Puebla98,Ostriker+99},
we can compute this necessary condition for BD formation as an integral
over the PDF of gas density. The result, discussed in \cite{Padoan+Nordlund04bd},
is that in typical molecular clouds approximately 1\% of the total 
mass is available for the formation of BDs. This fraction is approximately
10\% for conditions found in regions of cluster formation. 
Based on this result, the observed frequency of BDs is consistent
with the distribution of turbulent pressure, suggesting turbulence
may play an important role in the the origin of BDs.

In \cite{Padoan+Nordlund02IMF} the stellar IMF is interpreted
as the result of the turbulent fragmentation of star--forming 
clouds. It is shown that: i) The power law slope, $s$, of the 
stellar IMF is directly related to the turbulent power spectrum 
slope, $\beta$, $s=3/(4-\beta)$; ii) the abundance of stars of
mass smaller than the IMF peak depends on the probability density
function (PDF) of the gas density in the turbulent flow. 
As a result, the IMF is a power law with slope close to Salpeter's 
value \citep{Salpeter55} at large masses, and the abundance of 
low mass stars and BDs is a function of the turbulent rms 
velocity, the mean gas density and the mean temperature.
The IMF is fully determined once these physical parameters
of the star--forming region are known. 
As shown in Figure~\ref{f1}, the predicted BD abundance increases with
increasing Mach number and density (left panel) and is consistent
with the abundance inferred in stellar clusters (right panel).

\begin{figure*}[ht]
\centerline{
\epsfxsize=13.3cm \epsfbox{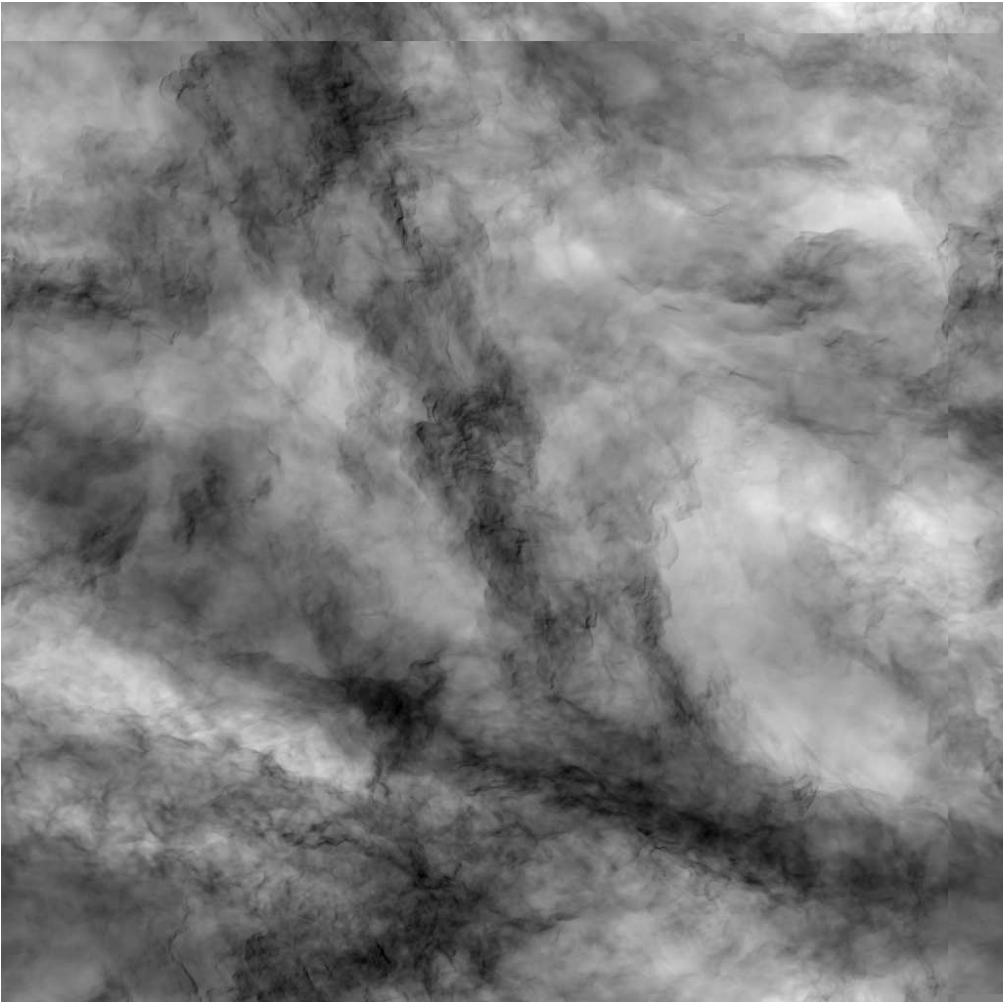}
}
\caption[]{Logarithm of the projected density field from an AMR 
simulation of isothermal, supersonic hydrodynamic turbulence with 
an effective resolution of $1024^3$ computational zones (Kritsuk, 
Padoan \& Norman, in preparation).}
\label{f2}
\end{figure*}

\section{AMR Experiments}

We employ a 3-D parallel structured adaptive mesh refinement (AMR)
code, {\em Enzo},\footnote{See {\tt http://cosmos.ucsd.edu/enzo/}}
developed at the Laboratory for Computational Astrophysics by Bryan,
Norman and collaborators \citep{bryan.99,oshea......04}. 
{\em Enzo} is a public domain Eulerian grid-based hybrid code which
includes hydrodynamic and N-body solvers and uses the AMR algorithm
of Berger \& Colella \citep{berger.89} to improve spatial resolution
in regions where that is required. The central idea behind AMR is
to solve the conservation laws on a grid, adding finer meshes in regions
that require enhanced resolution. AMR is spatially-- and time--adaptive,
can be used with accurate methods for solving the (magneto)hydrodynamic
equations. Mesh refinement can be automatically advanced to an arbitrary 
level, based on any combination of problem--specific local refinement 
criteria (e.g., density threshold, strong shocks, high
vorticity, large gradients, Jeans length \citep{truelove.....97}, 
cooling time, etc.). 

With this code, we have run the first AMR turbulence experiment 
ever attempted, achieving an effective resolution of $1024^3$
computational zones (Kritsuk et al. in preparation). 
The rms Mach number of the turbulent flow is $M_{\rm s}\approx 6$, 
and we have adopted an isothermal equation of state, periodic
boundary conditions, large scale random forcing and mesh refinement
based on the local shock or shear amplitude. The logarithm 
of the projected density field from a snapshot of this simulation 
is shown in Figure~\ref{f2}. The density field of this simulation
shows an incredibly rich structure, with sharp filaments on all 
scales, broken into even denser peaks. Some of these density
peaks are gravitationally unstable and would collapse if gravity
were included in the experiment. We refer to these density peaks as 
{\it turbulent seeds}, as they are the turbulent progenitors of
protostars. Turbulent seeds of BD mass are commonly found in
the turbulent flow. 

\begin{figure*}[ht]
\centerline{
\epsfxsize=13.7cm \epsfbox{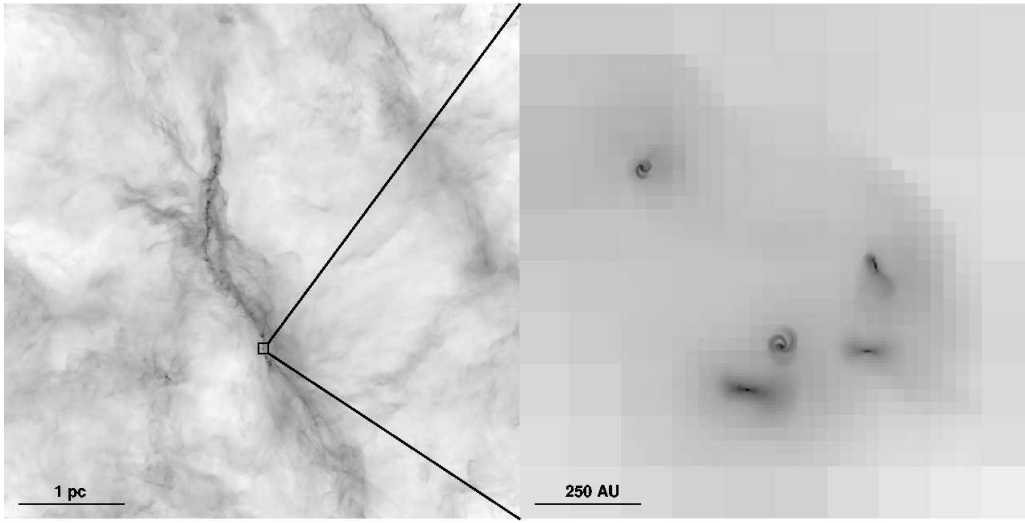}
}
\caption[]{Left panel: Logarithm of the projected gas density in 
our AMR simulation of star formation in a turbulent cloud (see text). 
Right panel: Higher resolution view of a small region inside the 
dense filament marked by the small square in the left panel. The 
small square is drawn larger than its actual size, as the magnification 
factor between the right and the left panels is 840.}
\label{f3}
\end{figure*}

In order to verify the outcome of the gravitational collapse of the 
turbulent seeds, we run a second experiment with the same code, where
self--gravity is included. In this experiment we use the AMR method 
to resolve the gravitational collapse. The computational box represents
a star--forming cloud of size 5~pc and average number density 
$500$~cm$^{-3}$. In the collapsing regions we are able to fully
resolve protostellar disks down to a scale of approximately 1~AU.
This simulation shows that the densest turbulent seeds, found 
primarily within dense filaments, collapse into protostars surrounded
by protostellar disks. Figure~\ref{f3} shows a typical filament 
containing many collapsing ``seeds'' (left panel). The seeds appear
as small dense cores, because of the limited dynamical rage of the
image. In fact, visualized with higher levels of refinement, they 
are seen to be collapsing into protostars with circumstellar disks
(right panel of Figure~\ref{f3}). Many of these protostars contain 
a fraction of a solar mass, some only a BD mass.

Further analysis and runs with longer integration time will allow  
to establish the protostellar mass distribution. However, these
simulations already teach us an important lesson: Turbulent seeds of BD
mass are common and may collapse into BDs, with no 
need to form BDs from the fragmentation of protostellar disks.
Simulations that do not resolve well the turbulent flow (like perhaps
every smooth particle hydrodynamic (SPH) simulation to date) are bound
to artificially suppress this important mechanism of BD formation.
Furthermore, insufficient numerical resolution may also cause 
artificial fragmentation of protostellar disks (Fisher et al., in 
preparation). Recent attempts to support the disk fragmentation origin 
of BDs, and even deny the importance of turbulent fragmentation, 
based on SPH simulations \citep{Bate+2002} are affected by both 
problems. We also warn against a direct application of results from 
simulations without magnetic fields to present day star formation, 
because magnetic fields affect the fragmentation mechanism 
\citep{Padoan+Nordlund99MHD} and the angular momentum transport.

\section{Conclusions}
High resolution numerical experiments of supersonic
turbulence and of star formation in turbulent clouds support the
suggestion that BDs are formed primarily from the process of turbulent
fragmentation, as {\it turbulent seeds}, like hydrogen burning stars.
As frequently pointed out at this workshop, all the observational
evidence points to a common origin of stars and BDs. Furthermore,
observations have shown that the mass distribution of
prestellar condensations is indistinguishable from the stellar IMF
\citep{Motte+98,Testi+Sargent98,Onishi+99,Johnstone+2000a,Johnstone+2000b,%
Motte+2001,Johnstone+2001,Onishi+2002}, both in the functional shape and in the
range of masses, including BD mass cores \citep{Walsh+04}. This direct
evidence of the turbulent seeds provides strong support to the
turbulent fragmentation origin of BDs.

\begin{acknowledgements}
The numerical simulations were done using the IBM Data Star system 
at the San Diego Supercomputer Center with support from NRAC award 
MCA098020S.

\end{acknowledgements}


\end{document}